\documentclass{article}
\usepackage{spconf,amsmath,graphicx}
\usepackage{amsmath}
\usepackage{amsmath,amssymb}
\usepackage{stmaryrd}
\usepackage{graphicx,graphics,color,psfrag}
\usepackage{cite,balance}
 \usepackage{subfigure}
\usepackage{caption}  
\allowdisplaybreaks
\usepackage{algorithm}
\usepackage{bm}
\usepackage{eufrak}
\usepackage{url}
\usepackage[english]{babel}
\usepackage{algpseudocode}
\usepackage{multirow}
\usepackage{enumerate}
\usepackage{cases}

\usepackage[utf8]{inputenc}
\usepackage[english]{babel}
\usepackage{amsthm}
 \usepackage{amsthm}

\newtheorem{proposition}{Proposition}

\title{Pushing The Limit of Type I Codebook For FDD Massive MIMO Beamforming: A Channel Covariance Reconstruction Approach}
\name{Kai Li$^{\sharp,\star}$, Ying Li$^{\sharp,\star}$, Lei Cheng$^{\star}$, Qingjiang Shi$^{\dagger, \star}$, and Zhi-Quan Luo$^{ \sharp,\star}$ \thanks{Kai Li and Ying Li are the co-first-authors (in alphabet order).}}
\address{$\star$ Shenzhen Research Institute of Big Data\\
$\sharp$ The Chinese University of Hong Kong, Shenzhen; $\dagger$ School of Software Engineering, Tongji University}
\begin{document}
%
\maketitle
\begin{abstract}
There is a fundamental trade-off between the channel representation resolution of codebooks and the overheads of feedback communications in the fifth generation new radio (5G NR) frequency division duplex (FDD) massive multiple-input and multiple-output (MIMO) systems. 
In particular, two types of codebooks (namely Type I and Type II codebooks) are introduced with different resolution and overhead requirements. Although the Type I codebook based scheme requires lower feedback overhead, its channel state information (CSI) reconstruction and beamforming performance are not as good as those from the Type II codebook based scheme. However, since the Type I codebook based scheme has been widely used in 4G systems for many years, replacing it by the Type II codebook based scheme overnight is too costly to be an option. Therefore, in this paper, using Type I codebook, we leverage advances in cutting plane method to optimize the CSI reconstruction at the base station (BS), in order to close the gap between these two codebook based beamforming schemes. Numerical results based on channel samples from QUAsi Deterministic RadIo channel GenerAtor (QuaDRiGa) are presented to show the excellent performance of the proposed algorithm in terms of beamforming vector acquisition. 
\end{abstract}
\section{Introduction}
\label{sec:intro}


Beamforming, a signal processing technique for directional signal transmission or reception, has become an integral part of current wireless communication standards in cellular systems (4G-LTE, 5G) and Wi-Fi networks 
(802.11 ac/ah) \cite{overview_bf}. More recently, with advanced beamforming technologies, 
 massive multiple-input multiple-output (MIMO), in which the base station (BS) is equipped with a large number of 
 antennas, has the potential to fully leverage the spatial freedoms offered by the large-scale antenna array to fulfill ever-increasing demands for both higher data-rate services from massive users and higher quality of service at each user-end \cite{massive_mimo}. Therefore, massive MIMO systems, together with advanced beamforming technologies, have been widely recognized as leading technologies in 5G wireless communications and beyond. 

However, the promise offered by beamforming in massive MIMO systems relies on the accurate channel state information (CSI) acquired at the BS \cite{nikos, rui, CE5, CE6, Lei}. Different from time division duplex (TDD) massive MIMO, CSI acquisition in frequency division duplex (FDD) massive MIMO systems is not a simple matter \cite{nikos, rui, CE5, CE6}. In the absence of channel reciprocity, the downlink CSI estimated at the receiver is usually quantized using a codebook and then is fed back to the BS through the control channel. This calls for a sophisticated codebook design to preserve the CSI as much as possible, and accordingly results in a trade-off between the resolution of channel representation and the overheads of feedback communications.

To adapt to diverse data services, 3GPP has introduced two types of codebooks (namely Type I and Type II codebooks) \cite{code1, code2},
which are with different channel representation resolution. 
In particular, with Type I codebook, implicit feedbacks at the user-end 
are utilized by sending a precoder matrix indicator (PMI) back to the BS,
which points to the index of a favourite codeword in a standardized codebook
\cite{code1}. Since only the PMI is fed back, relatively light overhead
can be achieved at the cost of sacrificing the CSI representation resolution. 
In contrast, the scheme using Type II codebook feedbacks both wideband and subband 
indices for a more accurate CSI representation \cite{code2}, while leading to much 
higher overheads of communications. System-level performance evaluations for these 
two codebook-based schemes can be found in \cite{code3, code4}, and it was reported
that Type II codebook based beamforming scheme could achieve up to $30\%$ performance
enhancement over the Type I codebook based counterpart. 

Even with excellent performance, the Type II codebook based scheme cannot replace the Type I codebook based counterpart overnight, which has been widely used in 4G communication systems for many years\cite{code1,code2}. Therefore, it calls for advances in optimizations to further enhance the beamforming performance of the Type I codebook based scheme, aiming at closing the gap between the two codebook based schemes. 
To achieve this, this paper proposes a novel CSI reconstruction algorithm by formulating the CSI reconstruction problem as a 
feasiblility problem and then solving it via the cutting plane method.
Notice that there is a closely related work \cite{nikos}, in which the transmission and feedback schemes are both significantly different from this work. These differences result in another optimization problem, which prohibits the straightforward extension of the approach in \cite{nikos}. 
Numerical results based on channel samples from QUAsi Deterministic RadIo channel GenerAtor (QuaDRiGa)\footnote{https://quadriga-channel-model.de.} are presented to demonstrate the performance of the proposed algorithm in terms of beamforming vector acquisition.

 
 The remainder of this paper is organized as follows. In Section II, the system model and problem formulation are introduced, base on which the CSI reconstruction algorithm is developed in Section III. In Section IV, numerical results are presented. Finally, conclusions are drawn in Section V.

\section{System Model And Problem Formulation}
\label{sec:pagestyle}
%

Consider a point-to-point MIMO link comprising an evolved node B (eNB) with $N_A$ antennas and a user equipment (UE) with $N_U$ antennas. Let $\mathbf H(t) \in \mathbb C^{N_U \times N_A}$ represent the downlink channel matrix between the eNB and the UE in time slot $t$. Assume that its covariance matrix\footnote{$\mathbb E [\mathbf H(t)]=0$.} $\mathbf C = \mathbb E[\mathbf H(t)^H \mathbf H(t)] \in \mathbb C^{N_A \times N_A}$ remains constant within $T$ consecutive time slots but may change afterwards. In each time slot, channel state information reference signal (CSI-RS) $\mathbf s(t) \in \mathbb C^{N_P \times 1}$ is utilized to help the UE acquire the CSI. A set of rank-1 codebook vectors $\mathcal V = \{\mathbf v_m \in \mathbb C^{N_P \times 1}, m = 1, \cdots, M\}$ is shared between the eNB and the UE. 
The number of the CSI-RS ports $N_P$ is usually smaller than the number of antennas $N_A$ at the eNB ( i.e., $N_P \leq N_A$). To tackle this, a pilot weighting matrix $\mathbf Q(t) \in \mathbb C^{N_A \times N_P}$ (also known as virtual antenna matrix) is adopted to enhance the dimension of the CSI-RS via linear transformation, i.e., $\mathbf Q(t) \mathbf s(t) \in \mathbb C^{N_A \times 1}$. Consequently, in each time slot, the received data at the UE can be modelled as 
 \begin{align}
 \mathbf y(t) = \mathbf H(t) \mathbf Q(t) \mathbf s(t) + \mathbf z(t),
 \end{align} 
 where $\mathbf z(t) \sim \mathcal {CN}(\mathbf z(t) | \mathbf 0, \sigma^2 \mathbf I_{N_U})$ is the additive white Gaussian noise (AWGN). From (1), it can be seen that the effective channel for the UE is 
 \begin{align}
 \mathbf H_e (t) = \mathbf H(t) \mathbf Q(t).
 \end{align}
 Therefore, without the knowledge of $\mathbf Q(t)$, the UE can only acquire the effective channel covariance matrix
 \begin{align}
 \mathbf R(t) = \mathbb E[ \mathbf H_e (t)^H \mathbf H_e (t)] = \mathbf Q(t)^H \mathbf C \mathbf Q(t),
 \end{align}
 and is assumed to obtain $\mathbf R(t)$ ideally. To feed the CSI back, directly transmitting $\mathbf R(t)$ from the UE to the eNB via the reverse link is however too communication/energy resource-demanding as an option. Alternatively, the best codebook index 
 \begin{align}
 m_0(t)&= {\arg\max}_{m=1,\cdots,M} \mathbf v_m^H \mathbf R(t) \mathbf v_m \nonumber\\
 & = {\arg\max}_{m=1,\cdots,M} \mathbf v_m^H \mathbf Q(t)^H \mathbf C \mathbf Q(t) \mathbf v_m,
 \end{align}
 and the channel quality indicator (CQI) value 
  \begin{align}
 \eta (t) =\mathbf v_{m_0(t)}^H \mathbf Q(t)^H \mathbf C \mathbf Q(t) \mathbf v_{m_0(t)}
 \end{align}
 are computed in the UE side and then are fed back to the eNB.

Using this scheme, although the communication overheads of the reverse link can be significantly reduced, a great challenge has been raised for the eNB to reconstruct the channel covariance matrix $\mathbf C$ from the collected codebook indices $\{m_0(t)\}_{t=1}^T$ and CQIs $\{\eta(t)\}_{t=1}^T$, which are up to the designed pilot weighting matrices $\{\mathbf Q(t)\}_{t=1}^T$. The coupling between $\{m_0(t), \eta(t)\}_{t=1}^T$ and $\{\mathbf Q(t)\}_{t=1}^T$ motivates the following problem formulation:
 \begin{align}
\min_{ \{\mathbf{Q}(t)\}_{t=1}^T} \mathrm{Volume}( \mathcal B ( \hat{\mathbf C}; \{\mathbf{Q}(t)\}_{t=1}^T)), \label{eq5}
\end{align}
where 
\begin{align}
	 &\mathcal B ( \hat{\mathbf C}; \{\mathbf{Q}(t)\}_{t=1}^T) \nonumber \\
	&= \Big \{\hat{\mathbf C } |~\mathbf {v}^H_{m} \mathbf Q(t)^H \hat{\mathbf C} \mathbf Q(t) \mathbf v_{m} \leq  \mathbf v^H_{m_0(t)} \mathbf Q(t)^H \hat{\mathbf C} \mathbf Q(t) \mathbf v_{m_0(t)}, \nonumber\\
	& ~~~~~~~~~~~~t = 1,\ldots, T,~~ \forall \mathbf v_m \in \mathcal V, \tag{C1} \\
	& ~~~~~~~~~~~~ \mathbf v_{m_0(t)}^H \mathbf Q(t)^H \hat{\mathbf C } \mathbf Q(t) \mathbf v_{m_0(t)} = \eta(t), \nonumber \\
	& ~~~~~~~~~~~~ t = 1,\ldots, T,~~ \tag{C2} \\ 
	& ~~~~~~~~~~~~\hat{\mathbf C } \succeq \mathbf 0_{N_A}, \tag{C3}\\
    & ~~~~~~~~~~~~\mathrm{Tr}(\hat{\mathbf C} ) \leq b \Big \}. \tag{C4} \label{eq6}
\end{align} 
In \eqref{eq5}, $\mathrm{Volume}(\cdot)$ measures the volume of set $\mathcal B ( \hat{\mathbf C};\!  \{\mathbf{Q}(t)\}_{t=1}^T)$. 
In $\mathcal B ( \hat{\mathbf C}; \{\mathbf{Q}(t)\}_{t=1}^T)$, the first set of inequalities (C1) defines a polyhedron $\mathcal P_t$ 
for time slot $t$, the constraint (C2) and (C3) confine the matrix $\hat{\mathbf C}$ to lie in hyper-planes and a positive semi-definite cone $\mathcal P_0$ respectively, 
and the last constraint (C4) is used to eliminate the scaling ambiguity. {The rationale behind problem \eqref{eq5} is that we hope to design weighting matrices $\{\mathbf Q(t)\}_{t=1}^T$ such that the set $\mathcal B ( \hat{\mathbf C}; \{\mathbf{Q}(t)\}_{t=1}^T)$, which is the feasible set of channel covariance matrices, can be very small. Then any point (e.g., the center) inside the set could give a good estimate of the covariance matrix $\mathbf C$.} However, how to design these weighting matrices $\{\mathbf Q(t)\}_{t=1}^T$ has not been investigated so far.

\section{Channel Reconstruction Algorithm Development}
\label{sec:typestyle}

 \subsection{Weighting Matrix Design} 
 
In order to minimize the volume of set $\mathcal B ( \hat{\mathbf C}; \{\mathbf{Q}(t)\}_{t=1}^T)$, the principle of weighting matrix design is to steer the intersection $ \mathcal P_0 \cap \mathcal P_1 \cap \mathcal P_2 \cdots \cap \mathcal P_T$ towards infinitesimal. In other word, polyhedron $\mathcal P_{t+1}$ should significantly cut the intersections of previous polyhedra. This principle motivates the adoption of ``neutral/ deep cut" idea in the literature of cutting plane methods\cite{cp4}. In particular, the next polyhedron $\mathcal P_{t+1}$ should deeply cut the feasible region $\mathcal B ( \hat{\mathbf C}; \{\mathbf{Q}(i)\}_{i=1}^t)$ in the sense that the set center $\hat {\mathbf C}(t)$ is excluded from (or on the supporting faces of) polyhedron $\mathcal P_{t+1}$. The problem can be stated as: 
\begin{align}
&\mathrm{Find}~ \mathbf Q(t+1) \nonumber \\
\mathrm{s.t.} ~~ & \mathbf{v}_{m^{\prime} (t+1) }^{H} \mathbf{Q}(t+1)^{H} \hat{\mathbf{C}}(t) \mathbf Q \mathbf{v}_{m^{\prime}(t+1)} \nonumber\\
& \geq \mathbf{v}^H_{m_0(t+1)} \mathbf{Q}(t+1)^{H} \hat{\mathbf{C}}(t) \mathbf{Q}(t+1) \mathbf{v}_{m_0(t+1)}, \nonumber\\
&\exists m^{\prime}(t+1) \in \{1,2,\cdots, M\}. \label{eq7}
\end{align}
The difficulty of problem \eqref{eq7} lies in that PMI $m_0(t+1)$ is unknown when designing $ \mathbf Q(t+1)$. Therefore, we need to design $ \mathbf Q(t+1) $ such that the inequality in problem \eqref{eq7} holds for all the possible choices of $m_0(t+1) \in \{1,2,\cdots,M \}$. To achieve this, we have the following propositions. 
\begin{proposition}
If $ \mathrm{rank} (\hat{\mathbf{C}}(t)) = N_A$, construct a matrix $\mathbf{R}(t+1)$ as follows:
\begin{align}
&\mathbf{R}(t+1) = \sigma_{1}(t+1) \mathbf{v}_{m^{\prime}(t+1)} \mathbf{v}_{m^{\prime}(t+1) }^{H}\nonumber\\
&+\ldots+\sigma_{N_{P}} (t+1)\mathbf{u}_{N_{P}-1}(t+1) \mathbf{u}_{N_{P}-1}(t+1)^{H},  \label{8}
\end{align}
where $m^{\prime}(t+1) \in \{1,2,\cdots,M\}$ ; hyper-parameters $\{\sigma_{n}(t+1)\}_{n=1}^{N_P}$ and $\{\mathbf{u}_{n}(t+1)\}_{n=1}^{N_P-1}$ are pre-selected such that $\sigma_{1}(t+1) \geq \sigma_{2}(t+1) \geq \cdots \geq \sigma_{N_{P}} (t+1)$ and column vectors $\{ \mathbf{v}_{m^{\prime}(t+1)},\mathbf{u}_{1}(t+1) ,\cdots, \mathbf{u}_{N_{P}-1}(t+1) \}$ all orthonormal.
The solution of problem (7) can be obtained via solving the equation $\mathbf{Q}^{H}(t+1)\hat{\mathbf{C}}(t)\mathbf{Q}(t+1) = \mathbf{R}(t+1)$, and takes the following expression:
\begin{align}
 \mathbf{Q}(t+1) = \hat{\mathbf{C}}(t)^{-\frac{1}{2}} \mathbf{X}(t+1) \mathbf{\Sigma}(t+1) \mathbf{Y}(t+1), \label{Q form 1}
\end{align}
where
\begin{align}
&\mathbf Y(t+1) \! = \!\left [ \mathbf{v}_{m^{\prime}(t+1)}, \!\mathbf{u}_{1}(t+1),\! \cdots,  \!\mathbf{u}_{N_{P}-1}(t+1) \right], \\
&\mathbf \Sigma(t+1) \nonumber\\
& = \mathrm{diag}\{ \underbrace{\sqrt{\sigma_{1}(t+1)}, \cdots, \sqrt{\sigma_{N_P}(t+1)}}_{N_P}, \underbrace{0, \cdots, 0}_{N_A-N_P} \},
\end{align}
and matrix $\mathbf{X}(t+1) \in \mathbb{C}^{N_{A} \times N_{A}}$ is a random unitary matrix, i.e., $\mathbf{X}(t+1) ^H\mathbf{X}(t+1) = \mathbf I_{N_A}$.
\end{proposition}

\begin{proposition}
	If $ \mathrm{rank} (\hat{\mathbf{C}}(t)) = K< N_A$, construct a matrix $\mathbf{R}(t+1)$ as follows:
	\begin{align}
	&\mathbf{R}(t+1) = \sigma_{1}(t+1) \mathbf{v}_{m^{\prime}(t+1)} \mathbf{v}_{m^{\prime}(t+1) }^{H} \nonumber\\
	&+\ldots+\sigma_{N_{P}} (t+1)\mathbf{u}_{N_{P}-1}(t+1) \mathbf{u}_{N_{P}-1}(t+1)^{H},  \label{12}
	\end{align}
	where $m^{\prime}(t+1) \in \{1,2,\cdots,M\}$ ; $\{\mathbf{u}_{n}(t+1)\}_{n=1}^{N_P-1}$ are pre-selected such that column vectors $\{ \mathbf{v}_{m^{\prime}(t+1)}, \mathbf{u}_{1}(t+1) ,\cdots, \mathbf{u}_{N_{P}-1}(t+1) \}$ all orthonormal. When $K \leq N_P$, the hyper-parameters $\{\sigma_{n}(t+1)\}_{n=1}^{N_P}$ follows 
\begin{align}
&\sigma_{1}(t+1) \geq \sigma_{2}(t+1) \geq \cdots \geq \sigma_{K}(t+1)  \nonumber\\
&\textgreater \sigma_{K+1} (t+1) = \sigma_{K+2} (t+1) = \cdots = \sigma_{N_P} (t+1) = 0. \nonumber
\end{align}
 However, when $N_P< K <N_A$, the values of $\{\sigma_{n}(t+1)\}_{n=1}^{N_P}$ need to satisfy $$\sigma_{1}(t+1) \geq \sigma_{2}(t+1) \geq \cdots \geq \sigma_{N_P} (t+1) > 0.$$
 The solution of problem (7) can be obtained via solving the equation $\mathbf{Q}^{H}(t+1)\hat{\mathbf{C}}(t)\mathbf{Q}(t+1) = \mathbf{R}(t+1)$, and takes the following expression:
	\begin{align}
	\mathbf{Q}(t+1) = \!
	&\left[
	\begin{array}{c}
		\!\!\![[ \mathbf{U}^{H}_{1}]_{(:, 1:K)}(t)]^{\!-1} \mathrm{diag}(\mathbf{v}(t))^{-1} \mathbf{U}^{H}_{1}(t)\mathbf{F}(t+1)\\ 
	\mathbf{0} 
	\end{array}
	\!\!\!\right ] \nonumber\\
	& + \mathbf{O}(t), \label{Q form 2}
	\end{align}
where 
	\begin{align}
	&\mathbf{F}(t+1)\in\mathbf{C}^{N_A \times N_P} \nonumber\\ &~~~~~~~~~~~~\text{ such that }~\mathbf{F}^{H}(t+1)\mathbf{F}(t+1)= \mathbf{R}(t+1);\\
	& \mathbf{U}^{H}(t) = 
	\left[
	\begin{array}{c}
	\mathbf{U}^{H}_1(t) \in \mathbb{C}^{K\times N_A} \\ \hline 
	\mathbf{U}^{H}_2(t) \in\mathbb{C}^{(N_A-K)\times N_A}
	\end{array}
	\right]\nonumber\\
	&~~~~~~~~~~~~ \text{ is eigenvector matrix of }\mathbf{\hat{C}}^{\frac{1}{2}}(t);\\
	& \mathbf{v} = \left[ \sigma^{c}_1(t), \cdots, \sigma^{c}_K(t) \right] \nonumber\\
	& ~~~~~~~~~~~~ \text{ contains nonzero sigular-values of }\mathbf{\hat{C}}^{\frac{1}{2}}(t);\\
	&\mathbf{O}(t) \in \mathrm{Null} (\mathbf{U}_{1}^H(t)).
	\end{align}
\end{proposition}
Due to the page limit, the proofs of Proposition 1 and Propositon 2 are not included in this paper. The basic idea of the proofs is to solve the equation $\mathbf{Q}(t+1)^{H} \hat{\mathbf{C}}(t) \mathbf{Q}(t+1)=\mathbf{R}(t+1)$ via linear algebra, where $\mathbf R(t+1)$ has its particular structure as specified in \eqref{8} or \eqref{12}. 

From {\bf Proposition 1} and {\bf Proposition 2}, the solution of problem \eqref{eq7} can be obtained in a closed-form. Notice that the solution is not unique, since it is up to different choices of $m^{\prime}(t+1) \in \{1,2,\cdots,M\}$, $\{\sigma_{n}(t+1)\}_{n=1}^{N_P}$, $\{\mathbf{u}_{n}(t+1)\}_{n=1}^{N_P-1}$ and unitary matrix $\mathbf{X}(t+1) $.

\subsection{Center Matrix Optimization}
In problem \eqref{eq5}, the center matrix of set $\mathcal B ( \hat{\mathbf C}; \{\mathbf{Q}(i)\}_{i=1}^t)$ needs to be provided. Motivated by analytical center optimization in cutting plane methods \cite{cp4, cp5}, given the weighting matrices $\{\mathbf{Q}(i)\}_{i=1}^t$, center matrix $\hat {\mathbf C}(t)$ can be sought via solving the following problem:
\begin{align}
&\max_{ {\mathbf C} } \sum_{i=1}^{t}\sum_{m=1}^M \frac{1}{\eta(i)} \log ( \mathbf v_{m_0(i)}^H \mathbf Q(i)^H {\mathbf C} \mathbf Q(i) \mathbf v_{m_0(i)} \nonumber\\
& - \mathbf v_m^H \mathbf Q(i)^H {\mathbf C} \mathbf Q(i) \mathbf v_m) + \log \det ( {\mathbf C} ) + \lambda ||{\mathbf C}||_{*}\nonumber \\
&~~\mathrm{s.t.} ~~ a \leq \mathrm{Tr}({\mathbf C} ) \leq b, \nonumber \\
&~~~~~~~~ \mathbf v^H_{m_0}(i) \mathbf Q(i)^H {\mathbf C} \mathbf Q(i) \mathbf v_{m_0}(i) = \eta (i), \label{eq12} \nonumber\\
&~~~~~~~~ i = 1,\ldots, T,
\end{align}
where $|| \cdot ||_{*}$ denotes the nuclear norm of the argument, and $\lambda$ is the regularization parameter. $a$ and $b$ are the lower bound value and the upper bound value of the trace of the ground-truth covariance $\mathbf C$, respectively. Notice that due to the incorporation of system model information, problem \eqref{eq12} is different from the standard formulation of finding an analytical center \cite{cp4, cp5}. In particular, a nuclear norm based regularization term is added to promote the low-rank structure of $\hat {\mathbf C}(t)$; CQI values are utilized to reweight each log barrier term; and further constraints are included to shrink the feasible region. Even with these differences, problem \eqref{eq12} is still convex and can be well solved by the CVX solver\footnote{http://cvxr.com/cvx/.}.

\subsection{Algorithm Summary}
From previous subsections, it is clear that the update of center matrix $\hat{\mathbf C}(t)$ relies on $\{\mathbf{Q}(i)\}_{i=1}^t$ while the update of weighting matrix $\mathbf {Q}(t)$ relies on $\hat{\mathbf C}(t-1)$. Therefore, the center matrix and the weighting matrix need to be alternately updated, giving rise to the iterative algorithm summarized in {\bf Algorithm 1} at the top of this page. Due to the page limit, the convergence property and computational complexity of the proposed algorithm are not included in this conference paper.

\begin{algorithm}[!t]
  \caption*{\bf Algorithm 1: Channel Covariance Reconstruction Algorithm}
\noindent {\bf Initialization:} Set initial value of $\mathbf {Q}(0)$. 

For the communication round $t$ ($t \geq 1$),

$~~$\noindent Update center matrix $\hat{\mathbf C}(t)$ via solving \eqref{eq12};

$~~$\noindent If $\mathrm{rank}\left(\hat{\mathbf C}(t)\right) = N_A$, update $\mathbf {Q}(t+1)$ using \eqref{Q form 1};

$~~$\noindent If $\mathrm{rank}\left(\hat{\mathbf C}(t)\right) < N_A$, update $\mathbf {Q}(t+1)$ using \eqref{Q form 2};

\noindent {Until the communication round $T$.}

\noindent {\bf Output:} Channel covariance estimation: $\hat{\mathbf C}(T)$.
\end{algorithm}

\section{Numerical Results and Discussions}
Consider a base station with $N_A = 32$ antennas and $N_P = 8$ ports, serving a user with $N_U = 2$ antennas. The ground-truth channel covariance matrix is generated by QuaDRiGa (the speed of user terminal is 3 km/h). The initial value of $\mathbf Q(0)$ follows the patent \cite{20}. The Type I codebook and Type II codebook are generated according to 5G NR standards \cite{21,22}. The ground-truth covariance matrix is normalized via Frobenius norm. In center matrix optimization, the upper bound value $b = 2$, the lower bound value $a = -\infty$ and $\lambda = 1$. In weighting matrix design, $m'(t+1)$ is randomly chosen from $\{1,\cdots, M\}$, $\{\sigma_{n}(t+1)\}_{n=1}^{N_P}$ are chosen to be $1$; and $\mathbf X(t+1)$ is randomly generated. Each point in the figure is obtained via averaging Monte-Carlo trials.

For FDD MIMO beamforming, the performance of CSI reconstruction is measured by the beam precision $p = \frac{\hat{\mathbf w}^H\mathbf C\hat{\mathbf w}}{d}$, where $\mathbf C$ is the ground truth channel covariance matrix, $\hat{\mathbf w}$ is the principal eigenvector of the reconstructed channel covariance $\hat{\mathbf C}$, and $d$ is the largest eigenvalue of $\mathbf C$\cite{20}. We present the beam precisions using Type I codebook and Type II
 codebook as the benchmarks. The beam precisions from the channel covariance reconstruction algorithm in patent \cite{20} are also provided as the benchmark. 
 
In Fig.1, beam precisions from different approaches versus communication rounds are presented. It is clear that the beam precision from Type II codebook is much higher than that of the Type I codebook. With Type I codebook, it can be seen that in the first 16 communication rounds, the proposed method is with lower beam precisions than those from the patent algorithm. However, after 17 communication rounds, the beam precision from the proposed method continues to increase and finally touches the Type II codebook precision after 33 communication rounds. This shows that the proposed channel reconstruction algorithm indeed helps the optimal beamforming vector acquisition and then closes the performance gap between these two codebook based schemes.

 \begin{figure}[!t]
	\centering
	\includegraphics[width= 3 in]{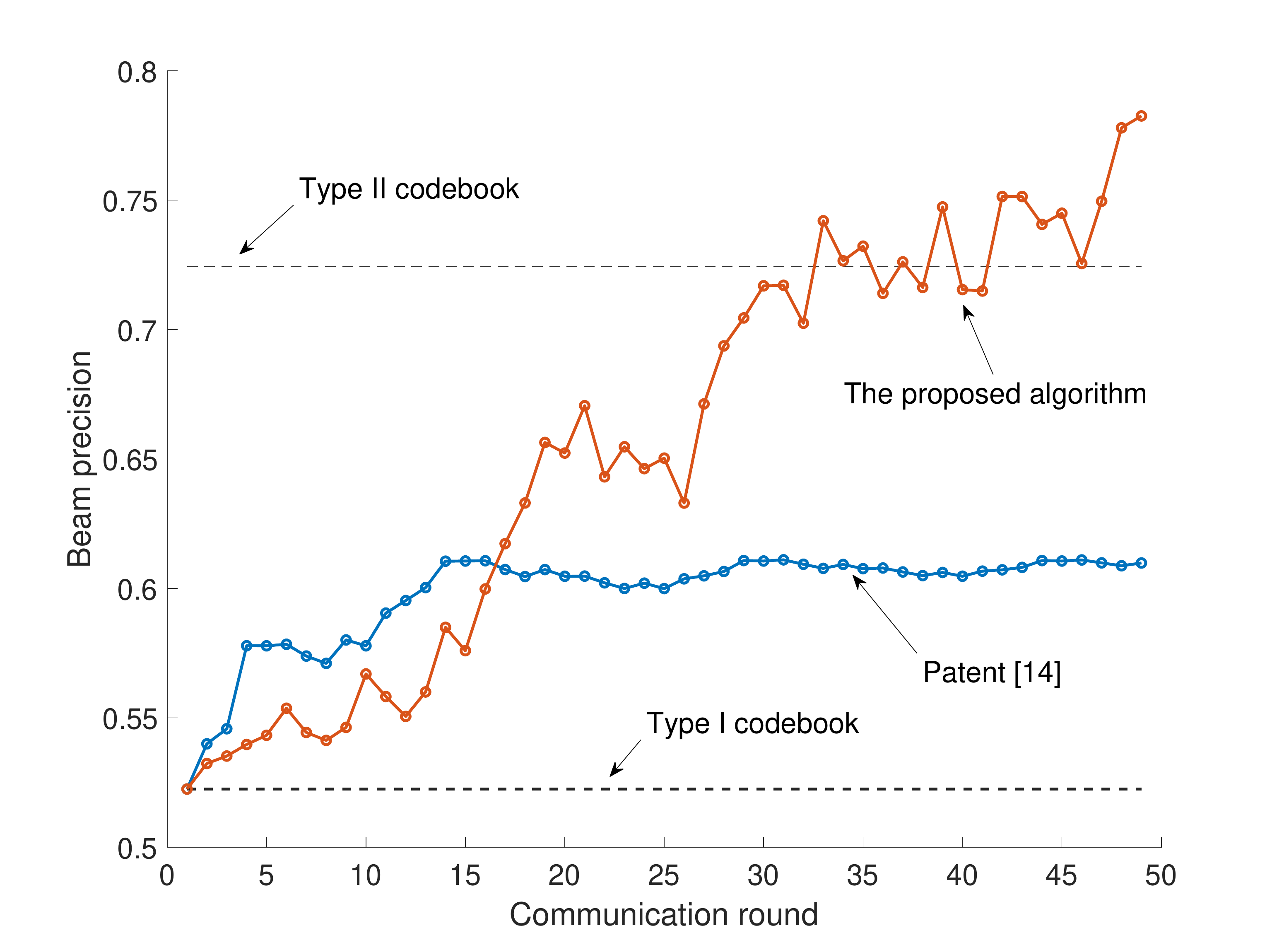}
	\caption{Beam precisions versus communication rounds.}
	\label{fig_topology}
 \end{figure}
 

\section{Conclusions}
In this paper, a channel covariance reconstruction algorithm was proposed for FDD massive MIMO communications with Type I codebook. Based on the reconstructed channel covariance matrix, the beamforming performance using Type I codebook can be significantly improved and even approach the counterpart using Type II codebook after several communication rounds. Numerical results have demonstrated the excellent performance of the proposed algorithm in terms of beamforming vector acquisition.

\vfill\pagebreak


\end{document}